\documentstyle[12pt]{article}

\def \be{\begin{equation}}
\def \ee{\end{equation}}
\def \bea{\begin{eqnarray}}
\def \eea{\end{eqnarray}}
\begin{document}
hep-th/9707060 
\begin{center}
{\Large{\bf{ Zamalodchikov's C-Theorem and The Logarithmic Conformal Field 
Theory
}}}
\vskip .5cm   
{\large{M.R. Rahimi-Tabar ${}^{2,3}$ and  S. Rouhani ${}^{1,2}$}}
\vskip .1cm
{\it{1) Sharif University of Technology, Tehran P. O. Box 11365-9161, Iran
\\2) Dept. of Physics , Iran  University of Science and Technology,\\
Narmak, Tehran 16844, Iran.
\\3) Institue for Studies in Theoretical Physics and 
Mathematics
\\ Tehran P.O.Box: 19395-5746, Iran.}}
\end{center}
\vskip .5cm
\begin{abstract}
 
We consider perturbation of a conformal field theory by a pair 
of relevant logarithmic operators and calculate the 
 beta function up to two loops. We observe that the beta function 
 can not be derived from a potential. Thus the renormalization
 group trajectories are not always along decreasing values of the 
 central charge. However there exists a domain of structure 
 constants in which the c-theorem still holds.
\end{abstract} 
\vskip .5cm                                                

PACS numbers, 11.25.Hf, 11.25.Db, 11.10.Gh    

{\bf 1- Introduction}

An important theorem was put forward by Zamalodchikov regarding the 
perturbation of conformally invariant field theories (CFT), which is 
known as $c$-theorem \cite {Zam}. The $c$-theorem states that there exists a 
function $C$,
of the coupling constants which is non-increasing along the trajectories of the 
renormalization group and its stationary points coincide with the fixed points
of the renormalization group. At these fixed points $C$ takes the value of the 
central charge.

On the other hand it is interesting to investigate the validity of the 
Zamalodchikov`s theorem beyond its original domain. One such group of 
theories are the logarithmic conformal field theories (LCFT). It has
been shown by Gurarie \cite {Gur} that CFT`s exist in 
which at least two primary fields have equal
conformal dimensions. 
Such a pair then have logarithms in their correlation functions.

The Logarithmic fields (operators) in CFT were first studied by Gurarie
in the $c=-2$ model \cite {Gur}. After Gurarie, 
these logarithms have been found 
in a multitude of other models such as the $WZNW$-model on the $GL(1,1)$ [3], 
the gravitationally dressed CFTs [4], $ c_{p,1}$ and non-minimal 
$c_{p,q}$ models [2,5-7], critical disordered models [8,9], and the $WZNW$ 
models
at level $0$ [10,11]. They play a role in the study of critical polymers
and percolation [5,6,12,13], 2D - turbulence [14-18] and quantum Hall states 
[19-21]. 
They are also important for studying the problem of recoil in 
the string theory and D-branes [10,22-25], as well as target space symmetries in
string theory [10]. The representation theory of the 
Virasoro algebra for LCFT was developed in [26]. The origin of the 
LCFT has been discussed in  [27,28,34].

Perturbing a stable fixed point by logarithmic 
operators has many repercussions. 
Firstly logarithmic operators come in at least pairs of relevant 
operators,
thus one always has to deal with a system of equations in renormalization 
flow trajectories.
Secondly the logarithmic response changes the renormalization flow.
Thirdly non-unitarity causes negative norms and this may affect the 
c theorem.

Unitarity is a prerequisite of the $c$-theorem and one may expect a 
break down of the $c$-theorem for non-unitary theories. 
Although most realizations of `s so far have been non-unitary, 
but unitary $LCFT$'s may also exist \cite{11}. 
Therefore c-theorem within the context of LCFT's is interesting from two 
different points of view; it may hold under certain conditions even for 
non-unitary LCFT's, it restricts the unitary LCFT's.

{ \bf 2-  The c - theorem} 

The proof of the $c$-theorem is based on conservation of the energy-
momentum
tensor and positivity, here we follow the proof given by Cardy \cite{Car}. 
In two
dimensions the energy-momentum tensor has three independent 
components, 
\bea
T = T_{zz} \hskip 1cm \bar T = T_{\bar z \bar z} \hskip 1cm U=T_{z 
\bar z} \eea

At a fixed point the theory is conformally invariant, the beta function 
vanishes and $ U = 0 $. Thus $T$ depends solely on $z$ and $\bar T$  
 on $\bar z$. The conservation of the energy-momentum tensor results in:
 
\bea
\partial_{\bar z} T + \frac  {1}{4} \partial_z U= 0 \cr
\partial_{z} \bar T + \frac  {1}{4} \partial_{\bar z} U= 0 
\eea

We are concerned with the perturbation of a fixed point Hamiltonian by an
operator $\Phi$:
\be
{\cal H = H_{*}} + g \Phi
\ee

The renormalization flow of the coupling constant $g$ is then given by 
calculating the change in the correlation functions of the theory 
perturbatively:
\be
< \cdots > = < \cdots >_* + g \int < \Phi \cdots >
\ee

We can now use the operator product expansion on the rhs. of eq.(4):
\be
T(z) \Phi (z_1) = \frac {h} {(z-z_1)^2} \Phi(z) + \frac {1-h} {z-z_1} 
\partial \Phi(z)
\ee

We observe that the rhs of eq.(4) is divergent thus needs regularization.
Consequently $\partial_z T$ no longer vanishes and we find:

\be
\partial_z T = -\pi (1-h) \partial_z \Phi
\ee

then the conservation of the energy-momentum tensor implies that:
\be
U=-4\pi g (1-h) \Phi
\ee

Here $U$ is the response of the action to the scale transformation 
$z\rightarrow \lambda z$. This is valid all the way along a trajectory leaving a 
fixed point until another fixed point is reached. At the second point the $UV$ 
behaviour changes.
 Let us consider the flow from a $UV$ fixed point, to a relatively $IR$
 fixed point. Given the spin structure of the three components of energy-
 momentum tensor, the following holds:
 \bea
 < T(z) T(0) > = \frac {F(z \bar z)} {z^4}  \cr
< \bar T(z) \bar T(0) > = \frac{G(z \bar z)} {z^4}  \cr
< U(z) U(0) > = \frac{H(z \bar z)} {z^2 {\bar z}^2}  
 \eea
 using the conservation of the energy-momentum tensor (i.e. eq.(2)), we have 
 
 \bea
 \dot F + \frac {1}{4} (\dot G - 3G)=0 \cr 
 \dot G - G + \frac {1}{4} (\dot H - 2H)=0
 \eea
 where $\dot F = z \bar z { F(z \bar z)}^{ \prime} $. Defining 
 $C=2F-G-\frac {3}{8}H$,
 we have
 \be
 \dot C=- \frac {3}{4} H
 \ee

Now in unitary theories we have $H>0$, thus $C$ is a non-decreasing function, 
and it is stationary only when $U$ is zero, that is at the conformally 
invariant points. Furthermore the quantities $G$ and $H$ vanish at the 
fixed point and $F=c/2$, thus we have $C=c$. This proof can be easily 
extended to the case of more than one operator.
\\
{\bf 3-  The Logarithmic Conformal Field Theories } 

In its simplest version a logarithmic field theory is characterized by a pair
of fields which mix due to a scale transformation:
\be
\Phi(z) \rightarrow \lambda^{-x} \Phi(z) 
\ee
\be
\Psi(z) \rightarrow \lambda^{-x} (\Psi(z) - \log (\lambda) \Phi )
\ee
Note that formally  one can think of $\Psi$ as derivative of $\Phi$ with
respect to $x$ \cite{28,sr,34}. The OPE with the energy momentum 
tensor likewise changes: 
\be
T(z) \Phi(z_1) = \frac { h \Phi} {(z-z_1)^2} + \frac {1-h} {(z - z_1)} 
\partial \Phi 
\ee
\be
T(z) \Psi(z_1) = \frac { h \Psi} {(z-z_1)^2} +  \frac {\Phi} {(z-z_1)^2} + 
\frac {1-h} {(z - z_1)} \partial \Psi - \frac {1} {(z-z_1)} \partial \Phi 
\ee
The consistency of invariance under the action of Virasoro algebra 
generators requires the two-point functions of $\Psi$ and $\Phi$ to 
have an unusual form:
\be
< \Phi (z) \Phi (0) > = 0
\ee
\be
< \Phi (z) \Psi (0) > = b z^{-2x} 
\ee
\be
< \Psi (z) \Psi (0) > = z^{-2x} ( d-2b \log (z) ) 
\ee
where $b$ and $d$ are constants. It is this above property which has an 
important bearing on the $c$-theorem. We perturb the fixed point 
Hamitonian ${\cal H_{*}}$ by a pair of operators $\Psi$
and $\Phi$ using two coupling constants $g_1$ and $g_2$:

\be
{\cal H= H_{*}} + \int d^2 z (g_1 \Phi + g_2 \Psi)
\ee
To make the coupling constants dimensionless, and also maintain the
invariance given by equations (11,12), we 
rewrite the above expression as follows:
\be
{\cal H= H_{*}} + \int d^2 z (g_2 a^{(x-2)} \Psi + G_1 a^{(x-2)} \Phi) 
\ee

where $a$ is the lattice constant and the coupling constant $G_1$ is :
\be
G_1=g_1 + \log(a) g_2
\ee 
 
The function $C( g_1 , g_2)$ now can be calculated using the 
two point functions and eq.(7) as: 
\be
C(g_1,g_2)=c^* - 6 \pi^2 (2 g_1 g_2 (2-x) b - g_2^2 b + g_2^2 (2-x)d)
\ee
Clearly $C$ does not always decrease as result of a change 
in scale, but it does if the following condition holds:

\be
2 ( 2-x) g_1 > (b - d (2-x)) g_2
\ee
This in turn holds if $x<2, g_1>0 , g_2>0 $ and;
\be
 b = d (2-x)
\ee
To calculate the renormalization flow we need  the OPE coefficients.
Up to two loops, the result of ordinary CFT still holds \cite{Car} 
provided we replace $g_1$ with $G_1$:
\be
\dot g_i=(2-x_i)g_i- \pi c_{i}^{jk} g_k g_j
\ee

Note that in LCFT the structure functions depend on $\log(r)$ through 
logarithmic terms and one has to take care when applying the above 
equation.
To proceed further we need the OPE of the fields $\Phi$ and $\Psi$ [ 28,34]:
\be
 \Phi (z) \Phi (0)  = \cdots + z^{-x} ( A - B \log (z) ) \Phi(0) + B 
z^{-x} \Psi(0)
\ee
\bea
 \Phi (z) \Psi (0) &=& \cdots + b z^{-2x} + z^{-x} (D -(A - E) \log(z)
  \cr &-& B (\log (z))^{2} ) \Phi(0) + z^{-x} (E-B \log (z))  \Psi(0)
 \eea 
\bea
 \Psi (z) \Psi (0) &=& \cdots + (d-2 b \log(z)) z^{-2x} + 
 z^{-x} ( G - (2D -K) \log(z)  \cr
&+& (A - 2E) (\log (z))^{2} + B \log(z)^{3} ) \Phi(0)  \cr
&+& z^{-x} (K-2E \log (z)+ B \log(z)^{2} )  \Psi(0)
 \eea 
here we have assumed that the 1-point functions of fields with zero 
conformal dimension vanishes, but this is need not be the case in 
nonunitary theories. The operator product expansion may change if 
this assumption is removed, and thus the renormalization flow will 
change. we shall deal with this case in a future work. 

After some algebra we obtain the renormalization flows:
\be
\dot g_1 = (2-x) g_1 - g_2 -\pi A g_1^2 - \pi G g_{2}^2 -2 \pi D g_1 
g_2 + \cdots
\ee
\be
\dot g_2 = (2 - x) g_2 - \pi K g_{2}^2 - 2 \pi E g_1 g_2 -\pi B g_1^2 
+\cdots
\ee
These equations clearly do not admit a potential, and even at the one 
loop approximation they have a Jordan form and cannot be 
diagonalized.

However the $UV$ fixed point ($g_1 = g_2 = 0$) is a stable point depending on 
whether $2-x$ is positive or not, in non-unitary theories, $x$ is negative thus 
the $UV$ point is always unstable. Although a potential does not 
exist it may still be that the flow minimizes some function such as $C$ 
as defined above. However a one loop calculation of $ C$ has indicated 
that $C$ is not always decreasing.

{\bf 4-  Discussion}

The above results can easily be generalized to the case where the 
Jordan cell has more than two members.
If the Hamiltonian is perturbed by more than two logarithmic fields:
\be
{\cal H = H_*} + \sum _{\alpha =1} ^N \Phi_{\alpha}
\ee
we then derive the following 1-loop equations:
\be
\dot g_N = (2-x) g_N
\ee
\be
\dot g_{N-1} = (2-x) g_{N-1} - g_N
\ee
and
\be
\dot g_1 = (2-x) g_1 - (N-1) g_2
\ee

To summarize we observe that the c-theorem does not always hold in logarithmic 
conformal field theories, but under certain conditions it may hold. When dealing  
with non-unitary theories this is not a disaster, but if we find 
unitary LCFT's this result will put a restriction on their structure 
constants. Some authors have discussed
the validity of the c-theorem in a wider context \cite {5,Kur,Cardy,Dam}
it may be that a different definition of C, is necessary to cover cases such as 
disorder.
We also observe that LCFT's can be formulated in terms of nilpotent 
parameters \cite{34} , we suspect that the above analysis should have 
a transparent form if expressed in this terminology, work in this 
direction is under progress.

\end{document}